\documentclass{article}
\usepackage{spconf,amsmath,graphicx}

\usepackage{subfigure} 
\graphicspath{{./image/}}

\usepackage{amssymb}

\DeclareMathOperator*{\softmax}{softmax}

\usepackage{epstopdf}
\usepackage{tabularx}
\usepackage{multirow}
\usepackage{textcomp}
\usepackage{url}
\usepackage{lipsum}
\usepackage{makecell}

\usepackage{bm}
\usepackage{calc}

\usepackage{color}
\usepackage{dsfont}

\title{Self-Attention Generative Adversarial Network \\ for Speech Enhancement}
\name{\begin{tabular}{c}Huy Phan$^{\ast 1}$, Huy Le Nguyen$^{2}$, Oliver Y. Ch\'{e}n$^{3}$, Philipp Koch$^{4}$, \\ Ngoc Q. K. Duong$^{5}$, Ian McLoughlin$^{6}$, Alfred Mertins$^{4}$ \end{tabular}}
\address{$^1$Queen Mary University of London, UK, {~~~~~}$^2$HCMC University of Technology, Vietnam \\
	$^3$University of Oxford, UK, {~~~~~} $^4$University of L\"ubeck, Germany \\
	$^5$InterDigital R\&D France, France, {~~~~~}  $^6$Singapore Institute of Technology, Singapore \\
	{$^\ast$Correspondence email: \tt h.phan@qmul.ac.uk} 
}

\begin{document}
	%\ninept
	%
	\maketitle
	\begin{abstract}
		Existing generative adversarial networks (GANs) for speech enhancement solely rely on the convolution operation, which may obscure temporal dependencies across the sequence input. To remedy this issue, we propose a self-attention layer adapted from non-local attention, coupled with the convolutional and deconvolutional layers of a speech enhancement GAN (SEGAN) using raw signal input. Further, we empirically study the effect of placing the self-attention layer at the (de)convolutional layers with varying layer indices as well as at all of them when memory allows. Our experiments show that introducing self-attention to SEGAN leads to consistent improvement across the objective evaluation metrics of enhancement performance. Furthermore, applying at different (de)convolutional layers does not significantly alter performance, suggesting that it can be conveniently applied at the highest-level (de)convolutional layer with the smallest memory overhead\footnote{\scriptsize Source code is available at \url{http://github.com/pquochuy/sasegan}}.
	\end{abstract}
	\vspace{-0.35cm}
	\section{Introduction}
	\label{sec:intro}
	\vspace{-0.25cm}
	
	Speech enhancement is useful in many applications, such as speech recognition \cite{Donahue2018, Xu2019, Weninger2015} and hearing aids \cite{Schasse2014,Yang2005}. Recently, the research community has witnessed a shift in methodology from conventional signal processing methods \cite{Gerkmann2011,Ephraim1985} to data-driven enhancement approaches, particularly those based on deep learning paradigms \cite{Xu2015, Li2018, Weninger2015, Park2017,Erdogan2015}. Beside discriminative modeling with typical deep network variants, such as deep neural networks (DNNs) \cite{Xu2015}, convolutional neural networks (CNNs) \cite{Li2018, Park2017}, and recurrent neural networks (RNNs) \cite{Erdogan2015, Weninger2015}, generative modeling with GANs \cite{Goodfellow2014} have been shown to hold promise for speech enhancement \cite{Pascual2017, Zhang2020, Phan2020}. Furthermore, the study in \cite{Phan2020} indicates that generative modeling with GANs may result in fewer artefacts than discriminative methods.
	
	Since the seminal work \cite{Pascual2017}, SEGAN has been improved in various ways. Different input types have been exploited, e.g. raw waveform \cite{Phan2020,Baby2019} and time-frequency image \cite{Li2018,Zhang2020}. Better losses, like Wasserstein loss \cite{Zhang2020}, relativistic loss \cite{Baby2019}, and metric loss \cite{Zhang2020}, have been tailored to gain stabilization in the training process. SEGANs that learn multi-stage enhancement mappings have also been proposed \cite{Phan2020}. However, convolutional layers are still, and will probably remain, the backbone of these SEGAN variants. This reliance on the convolution operator limits SEGAN's capability in capturing long-range dependencies across an input sequence due to the convolution operator's local receptive field. Temporal dependency modeling is, in general, an integral part of a speech modeling system \cite{Defossez2020, Pham2019}, including speech enhancement when input is a long segment of signal with a rich underlying structure. However, it has mostly remained uncharted in SEGAN systems.

	On the one hand, self-attention has been successfully used for sequential modeling in different speech modeling tasks \cite{Pham2019, Sperber2018, Tian2019}. On the other hand, it is more flexible in modeling both long-range and local dependencies and is more efficient than RNN \cite{Zhang2020} in terms of computational cost, especially when applied to long sequences. The reason is that RNN is based on temporal iterations which cannot be parallelized whereas self-attention is based on matrix multiplication which is highly parallelizable and easily accelerated. We, therefore, propose a self-attention layer following the principle of non-local attention \cite{Wang2018, Zhang2019} and couple it with the (de)convolutional layers of a SEGAN to construct a self-attention SEGAN (SASEGAN for short). We further conduct analysis of how the proposed self-attention layer applied at different (de)convolutional layers will affect the enhancement performance. We will show, when equipped with a sequential modeling capability, the proposed SASEGAN leads to better enhancement performance than the SEGAN baseline across all the objective evaluation metrics. Furthermore, the performance gain is consistent regardless of which (de)convolutional layer the self-attention is applied, allowing it to be integrated into a SEGAN with a very small additional memory footprint.
	
	\vspace{-0.2cm}
	\section{Self-attention SEGAN}
	\vspace{-0.2cm}
	\subsection{SEGAN}
\begin{figure} [!t]
		\centering
		\includegraphics[width=0.75\linewidth]{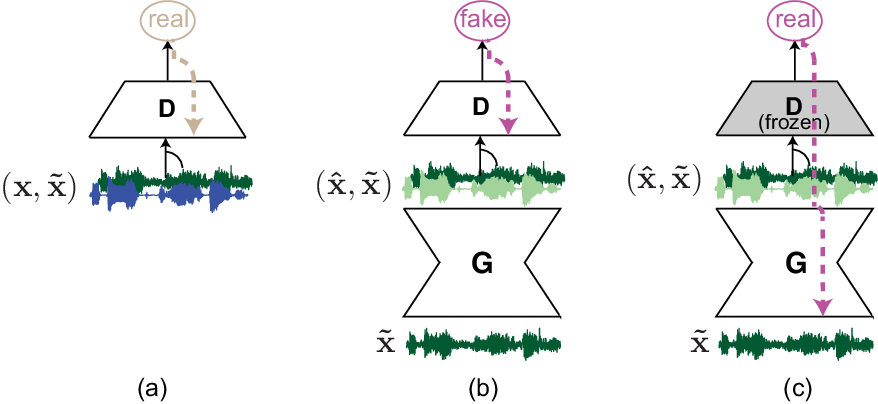}
		\vspace{-0.3cm}
		\caption{\small Adversarial training of GAN-based speech enhancement methods: $D$ learns to classify the pair $(\mathbf{x}, \mathbf{\tilde{x}})$ as real (a), and $(\mathbf{\hat{x}}, \mathbf{\tilde{x}})$ as fake (b). $G$ learns to fool $D$ so that $D$ classifies $(\mathbf{\hat{x}}, \mathbf{\tilde{x}})$ as real (c). Dashed lines represent the flow of gradient backpropagation.}
		\label{fig:segan_learning}
		\vspace{-0.1cm}
	\end{figure}

	\vspace{-0.15cm}
	Given a noise-corrupted raw audio signal $\mathbf{\tilde{x}}\!=\!\mathbf{x}\!+\!\mathbf{n}\!\in\!\mathbb{R}^T $, where $\mathbf{x} \in \mathbb{R}^T$ denotes a clean signal and $\mathbf{n}\!\in\!\mathbb{R}^T$ denotes additive background noise, the goal of speech enhancement is to find a mapping $f(\mathbf{\tilde{x}}):\mathbf{\tilde{x}}\mapsto\!\mathbf{x}$ to recover the clean signal $\mathbf{x}$ from the noisy signal $\mathbf{\tilde{x}}$. SEGAN methods \cite{Pascual2017, Zhang2020, Phan2020} achieve this goal by designating the generator $G$ as the enhancement mapping, i.e. $\mathbf{\hat{x}}\!=\!G(\mathbf{z}, \mathbf{\tilde{x}})$ where $\mathbf{z}$ is a latent variable. The discriminator $D$ is tasked to distinguish the enhanced output $\mathbf{\hat{x}}$ from the real clean signal $\mathbf{x}$. To this end, $D$ learns to classify the pair $(\mathbf{x}, \mathbf{\tilde{x}})$ as real and $(\mathbf{\hat{x}}, \mathbf{\tilde{x}})$ as fake. At the same time, $G$ learns to produce as good an enhanced signal $\mathbf{\tilde{x}}$ as possible to fool $D$ such that $D$ classifies $(\mathbf{\hat{x}}, \mathbf{\tilde{x}})$ as real. 
	%SEGAN is trained in this adversarial manner, as illustrated in Fig. \ref{fig:segan_learning}, to the objective function:
	SEGAN is trained in this adversarial manner, as illustrated in Fig.~\ref{fig:segan_learning}.
%	\begin{align}
%	\min_G\max_D V\!(D,G) = \mathbb{E}_{\mathbf{x},\mathbf{\tilde{x}}\sim p_{\text{data}}(\mathbf{x},\mathbf{\tilde{x}})}\!\log\!D(\mathbf{x},\mathbf{\tilde{x}}) \nonumber \\ +\,\mathbb{E}_{\mathbf{z}\sim p_{\mathbf{z}}(\mathbf{z}),\mathbf{\tilde{x}}\sim p_{\text{data}}(\mathbf{\tilde{x}})}\!\log(1 - D(G(\mathbf{z},\mathbf{\tilde{x}}), \mathbf{\tilde{x}})).
%	\end{align}
%	over the training set of $N$ pairs $\mathcal{X}=\{(\mathbf{x}, \mathbf{\tilde{x}})\}^N_{n=1}$.	
	Various losses have been proposed to improve adversarial training, such as least-squares loss \cite{Pascual2017,Mao2017}, Wasserstein loss \cite{Zhang2020}, relativistic loss \cite{Baby2019}, and metric loss, \cite{Zhang2020}. Here, we employ the least-squares loss as in the seminal work \cite{Pascual2017}. The least-squares objective functions of $D$ and $G$ are explicitly written as
	\begin{align}
	\small
	\min_D\mathcal{L}_{\text{LS}}(D) = &\frac{1}{2}\mathbb{E}_{\mathbf{x},\mathbf{\tilde{x}}\sim p_{\text{data}}(\mathbf{x},\mathbf{\tilde{x}})}(D(\mathbf{x},\mathbf{\tilde{x}}) - 1)^2 \nonumber \\ &+ \frac{1}{2}\mathbb{E}_{\mathbf{z}\sim p_{\mathbf{z}}(\mathbf{z}),\mathbf{\tilde{x}}\sim p_{\text{data}}(\mathbf{\tilde{x}})}D(G(\mathbf{z},\mathbf{\tilde{x}}), \mathbf{\tilde{x}})^2,
	\label{eq:D_objective} \\
	\min_G\mathcal{L}_{\text{LS}}(G)=&\frac{1}{2}\mathbb{E}_{\mathbf{z}\sim p_{\mathbf{z}}(\mathbf{z}),\mathbf{\tilde{x}}\sim p_{\text{data}}(\mathbf{\tilde{x}})}(D(G(\mathbf{z},\mathbf{\tilde{x}}), \mathbf{\tilde{x}})-1)^2 \nonumber\\ &+ \lambda|| G(\mathbf{z},\mathbf{\tilde{x}}) - \mathbf{x}||_1.
	\label{eq:G_objective}
	\end{align}

	\vspace{-0.4cm}
	\subsection{Self-attention SEGAN (SASEGAN)}
	\vspace{-0.1cm}
	\subsubsection{Self-attention layer}
	\vspace{-0.1cm}
	\label{sssec:selfattention}
	The proposed self-attention layer is adapted from the non-local attention \cite{Wang2018, Zhang2019}. Given the feature map $\mathbf{F}\!\in\!\mathbb{R}^{L\times C}$ output by a convolutional layer, where $L$ is the time dimension, $C$ is the number of channels. Note that the feature dimension is one since we are using 1D convolution to deal with raw speech input in this case. The query matrix $\mathbf{Q}$, the key matrix $\mathbf{K}$, and the value matrix $\mathbf{V}$ are obtained via transformations:
	\begin{align}
	\mathbf{Q} = \mathbf{F}{\mathbf{W}_Q} \text{,~~} \mathbf{K} = \mathbf{F}{\mathbf{W}_K} \text{,~~} \mathbf{V} = \mathbf{F}{\mathbf{W}_V},
	\end{align}
	where $\mathbf{W}_Q \in \mathbb{R}^{C \times \frac{C}{k}}$, $\mathbf{W}_K \in \mathbb{R}^{C \times \frac{C}{k}}$, and $\mathbf{W}_V \in \mathbb{R}^{C \times \frac{C}{k}}$ denote the weight matrices which are implemented by a $1\times1$ convolution layer of $\frac{C}{k}$ filters. That is, in the new feature spaces, the channel dimension is reduced by the factor $k$ mainly for memory reduction. Furthermore, given the $O(n^2)$ memory complexity, we also reduce the number of keys and values (i.e.~the time dimension of $\mathbf{K}$ and $\mathbf{V}$) by a factor of $p$ for memory efficiency. This is accomplished by a max pooling layer with filter width %**IVM deleted: of $p$  
and stride size of $p$. We use $k=8$ and $p=4$ here. The size of the matrices are, therefore, $\mathbf{Q} \in \mathbb{R}^{L \times \frac{C}{k}}$, $\mathbf{K} \in \mathbb{R}^{\frac{L}{p} \times \frac{C}{k}}$, and $\mathbf{V} \in \mathbb{R}^{\frac{L}{p} \times \frac{C}{k}}$. 
	The attention map $\mathbf{A}$ and the attentive output $\mathbf{O}$ are then computed as
	\begin{align}
	\mathbf{A} &= \softmax(\mathbf{\mathbf{Q}\mathbf{\bar{K}}^\mathsf{T}}), \text{~~} \mathbf{A} \in \mathbb{R}^{L \times \frac{L}{p}}, \\
	\mathbf{O} &= (\mathbf{A}\mathbf{V})\mathbf{W}_O, \text{~~} \mathbf{W}_O \in \mathbb{R}^{\frac{C}{k} \times C}.
	\end{align}	
	Each element $a_{ij} \in \mathbf{A}$ indicates the extent to which the model attends to the $j^{th}$ column $\mathbf{v}_j$ of $\mathbf{V}$ when producing the $i^{th}$ output $\mathbf{o}_i$ of $\mathbf{O}$. In addition, a transformation with weight $\mathbf{W}_O$ realized by a $1\times 1$ convolution layer of $C$ filters is applied to $\mathbf{A}\mathbf{V}$ to restore the shape of $\mathbf{O}$ to the original shape $L \times C$.

\begin{figure} [!t]
	\centering
	\includegraphics[width=0.725\linewidth]{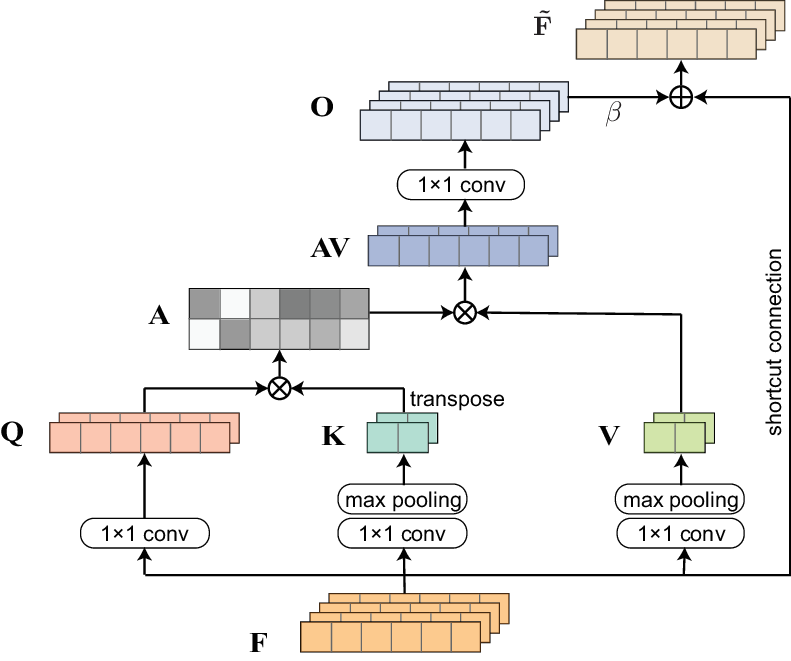}
	\vspace{-0.35cm}
	\caption{\small Illustration of the processing steps of the proposed self-attention layer with $L = 6$, $C = 4$, $p = 3$, and $k=2$.}
	\label{fig:selfattention}
	\vspace{-0.1cm}
\end{figure}

	Finally, we make use of a shortcut connection to facilitate information propagation, with the final output given as:
	\begin{align}
	\mathbf{\tilde{F}} = \beta\mathbf{O} + \mathbf{F},
	\end{align}	
	where $\beta$ is a learnable parameter. We illustrate the processing steps of a simplified self-attention layer with $L = 6$, $C = 4$, $p = 3$, and $k=2$ in Fig.~\ref{fig:selfattention}.

\vspace{-0.2cm}
\subsubsection{Network architecture}
\begin{figure} [!t]
	\centering
	\includegraphics[width=0.75\linewidth]{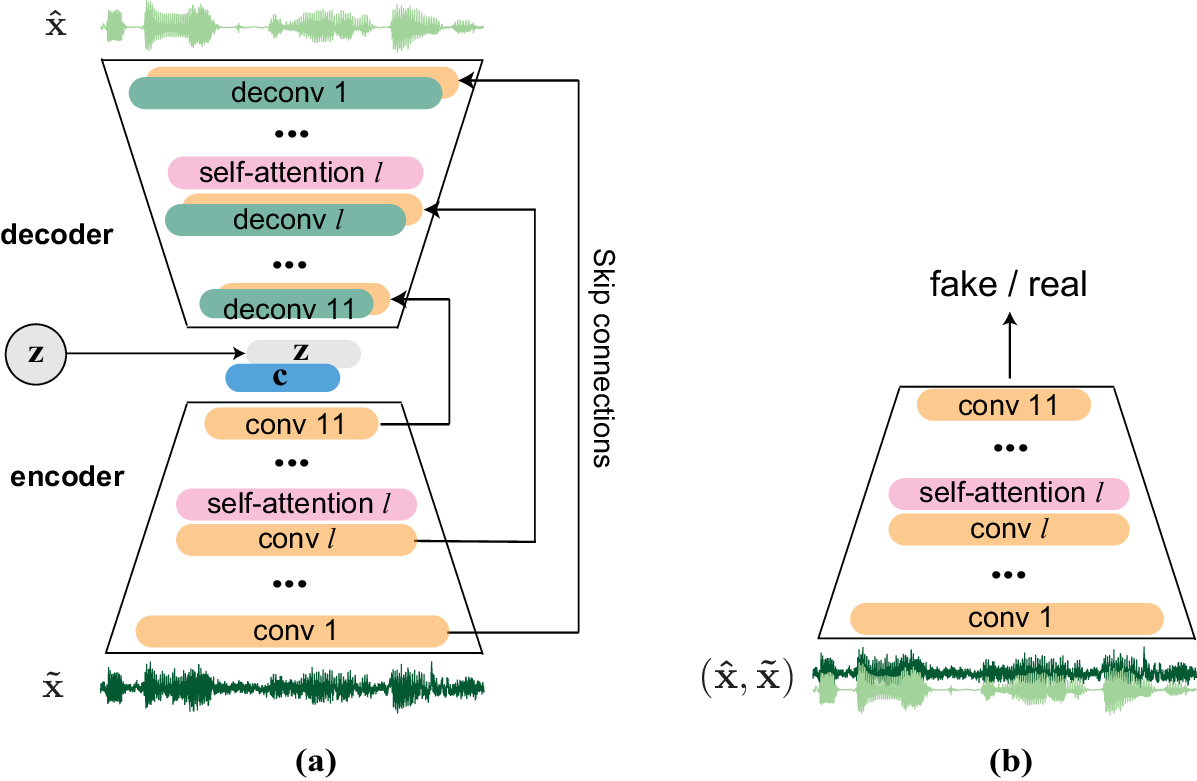}
	\vspace{-0.4cm}
	\caption{\small Illustration of SASEGAN. (a) the generator, (b) the discriminator.}
	\label{fig:sasegan}
	\vspace{-0.35cm}
\end{figure}

\vspace{-0.1cm}
Similar to SEGAN, the generator receives a raw-signal input of length $L\!\!=\!\!16,384$ samples (approximately one second at 16 kHz) and features an encoder-decoder architecture with fully-convolutional layers \cite{Radford2016}, as illustrated in Fig.~\ref{fig:sasegan}\,(a).  The encoder consists of 11 one-dimensional strided convolutional layers with a common filter width of 31, a stride of 2, and increasing number of filters $\{16, 32, 32, 64, 64,$ $128, 128, 256, 256, 512, 1024\}$, resulting in feature maps of size $8192\!\times\!16$, $4096\!\times\!32$, $2048\!\times\!32$, $1024\!\times\!64$, $512\!\times\!64$, $256\!\times\!128$, $128\!\times\!128$, $64\!\times\!256$, $32\!\times\!256$, $16\!\times\!512$, $8\!\times\!1024$, respectively. The noise sample $\mathbf{z} \in \mathbb{R}^{8 \times 1024}$ is then stacked on the last feature map and presented to the decoder. The decoder, on the other hand, mirrors the encoder architecture to reverse the encoding process by means of deconvolutions. All the (de)convoltuional layers are followed by parametric rectified linear units (PReLUs) \cite{He2015}.
%(i.e. fractional-strided transposed convolution). 
In order to allow information from the encoding stage to flow into the decoding stage, a skip connection is used to connect each convolutional layer in the encoder to its mirrored deconvolutional layer in the decoder (cf.~Fig.~\ref{fig:sasegan}\,(a)).

%---

%**IVM changed figure references in the text in the  paras above and below
The discriminator architecture, as illustrated in Fig.~\ref{fig:sasegan}\,(b), is similar to the encoder part of the generator. However, it receives a pair of raw audio segments as input. Its convolutional layers are also associated with virtual batch-norm \cite{Salimans2016} and Leaky ReLU activation \cite{Maas2013} with $\alpha=0.3$. The last convolutional feature map of size $8 \times 1024$ is further processed by an additional $1\times1$ convolutional layer and reduced to 8 features which are used for classification with softmax.

SASEGAN couples the self-attention layer described in Section \ref{sssec:selfattention} with the (de)convolutional layers of both the generator and the discriminator. Fig.~\ref{fig:sasegan}\,(a) and (b) show an example where the self-attention layer is coupled with the $l^{th}$ (de)convolutional layer. In general, the self-attention layer can be used in combination with any number, even all, of the (de)convolutional layers. We will investigate the self-attention layer placement in Section \ref{sec:experiments}. In SASEGAN, spectral normalization \cite{Miyato2018} is applied to all the (de)convolutional layers of the generator and the discriminator.

	\vspace{-0.3cm}
	\section{Experiments}
	\label{sec:experiments}
	\vspace{-0.3cm}
	\subsection{Experimental setup}
	\vspace{-0.2cm}
	We set two objectives in the conducted experiments. First, we studied and quantified the effects of using the proposed self-attention layer in speech enhancement. Second, we aimed to analyze the influence of self-attention layer placement in the generator and the discriminator, on enhancement performance. For the former, we used the SEGAN (i.e.~without self-attention) \cite{Pascual2017} as the baseline for comparison. For the latter, we evaluated SASEGAN with different values of the (de)convolutional layer index $l \in \{3, 4, \ldots, 11\}$. Note that, we were unable to experiment with the early (de)convolutional layers (i.e.~$1^{st}$ and $2^{nd}$) due to GPU memory limitations given the large time dimension of their feature maps, $8192$  and $4096$, respectively. We also studied the case when the self-attention layer was combined with all of the $l^{th}$ (de)convolutional layers where $l \in \{3, 4, \ldots, 11\}$.

	\vspace{-0.31cm}
	\subsection{Dataset}
	\vspace{-0.21cm}
	 The experiments were based on the database introduced in \cite{Botinhao2016}. This is also the one used in \cite{Pascual2017} to evaluate the SEGAN baseline. It consists of data from 30 speakers extracted from the Voice Bank corpus \cite{Veaux2013}. Ten types of noise 
	 %(two artificial and eight stemmed from the Demand database \cite{Thiemann2013})
	  were combined at signal-to-noise ratios (SNRs) of 15, 10, 5, and 0 dB, to introduce 40 noisy conditions to the training data. Similarly, 20 noisy conditions were introduced to the test data by combining five types of noise from the Demand database \cite{Thiemann2013} with four SNRs (17.5, 12.5, 7.5, and 2.5 dB). This resulted in 10 and 20 utterances for each noise condition per speaker in the training and test data, respectively. Adhering to prior works~\cite{Pascual2017, Phan2020, Zhang2020}, data from 28 speakers was used for training and data from two remaining speakers was used for testing. All audio signals were downsampled to 16 kHz. 
	\vspace{-0.31cm}
	\subsection{Parameters}
	\vspace{-0.21cm}
	The implementation was based on the Tensorflow framework \cite{Abadi2016}. Networks were trained with RMSprop \cite{Tieleman2012} for 100 epochs with a minibatch size of 50. During training, raw speech segments (of length 16,384 samples each) in a batch were sampled from the training utterances with 50\% overlap, followed by a high-frequency preemphasis filter with the coefficient of $0.95$. The trained network was then applied to the test utterances for enhancement purpose. For each utterance, raw speech segments were extracted without overlap, processed by the trained network, deemphasized, and concatenated to result in the enhanced utterance.

	\vspace{-0.31cm}
	\subsection{Experimental results}
	\vspace{-0.21cm}
	We used five objective signal-quality metrics: PESQ (in range $[-0.5, 4.5]$), CSIG (in range $[1,5]$), CBAK (in range $[1,5]$), COVL (in range $[1,5]$), and SSNR (in range $[1,\infty$]); and the speech intelligibility measure STOI (\%) \cite{Taal2011} for evaluation. As in \cite{Phan2020},  the five latest network snapshots were used and the results were averaged over 824 utterances of the test data.
	
	The  results obtained by the proposed SASEGAN alongside the SEGAN baseline and the noisy speech signals (without enhancement) are shown in Table~\ref{tab:results}. Note that in the table, we denote the SASEGAN with self-attention at the $l^{th}$ (de)convolutional layer as SASEGAN-$l$, $3\!\le\!l\!\le\!11$, and the one with self-attention at all the $l^{th}$ (de)convolutional layers, where $3\!\le\!l\!\le\!11$, as SASEGAN-All. Overall, introducing self-attention to the SASEGANs led to consistent improvements over the SEGAN baseline across all the objective metrics. Averaging over the SASEGAN-$l$s, absolute gains of $0.15$, $0.13$, $0.14$, $0.15$, $0.69$, and $0.2$ were obtained on PESQ, CSIG, CBAK, COVL, SSNR, and STOI over the SEGAN baseline, respectively. The performance was further boosted, although modestly, from the average when multiple self-attention layers were employed in SASEGAN-All, with absolute gains of $0.17$, $0.15$, $0.18$, $0.17$, $0.91$, and $0.37$, respectively. These gains, however, were achieved at the cost of increased computation time and memory requirements. 

\begin{figure} [!t]
	\centering
	\includegraphics[width=0.85\linewidth]{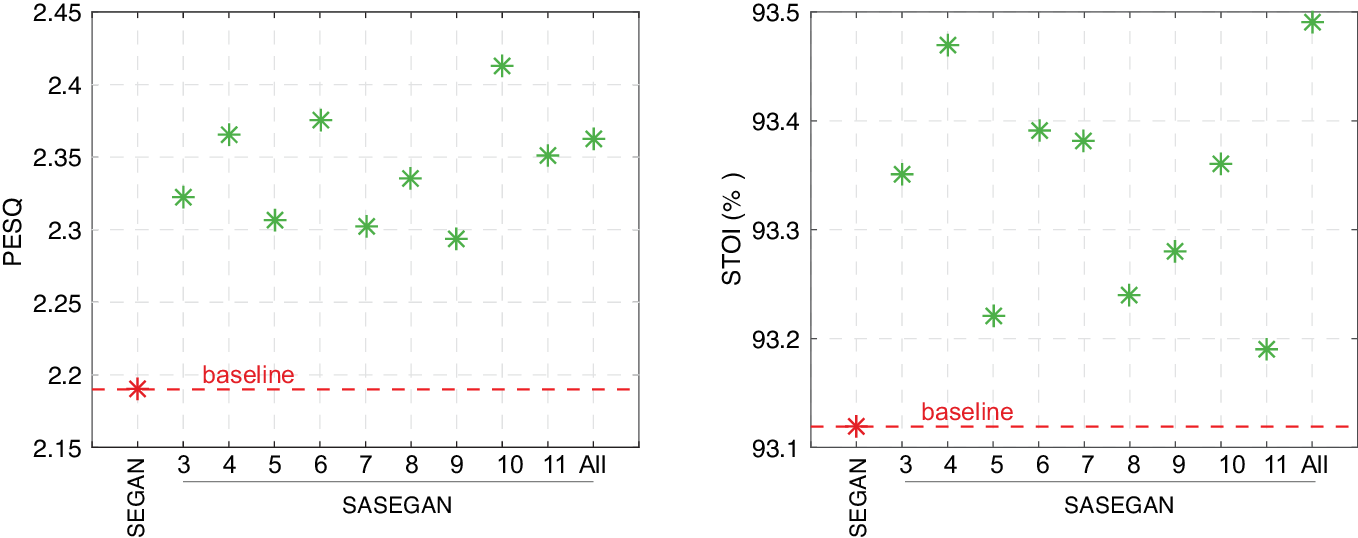}
	\vspace{-0.3cm}
	\caption{\small PESQ and STOI gains obtained by the SASEGAN-$l$s, $3 \le l \le 11$ over the SEGAN baseline.}
	\label{fig:pesq_stoi}
	\vspace{-0.2cm}
\end{figure}

	 Furthermore, using self-attention at different layer indices $l$ did not show a clear difference between the performance improvements of SASEGAN-$l$s over the SEGAN baseline, as depicted in Fig.~\ref{fig:pesq_stoi} for PESQ and STOI. This suggests that self-attention applied to the high-level (de)convolutional layer is expected to be as good as when applied in a low-level (de)convolutional layer. More importantly, by doing so, extra memory requirements are exponentially reduced and can be as little as $8^2 = 64$ memory units at $l=11$ given that the time dimension of the feature map at the $l^{th}$ (de)convolutional layer is given by $\frac{L}{2^l}$.
	 
	 \setlength\tabcolsep{1.25pt}
	 \begin{table}[b!]
	 	\vspace{-0.4cm}
	 	\caption{\small Results obtained by the studied speech enhancement systems  on the objective evaluation metrics. We highlight in bold where the proposed SASEGAN outperforms the baseline SEGAN.}
	 	%\small
	 	\vspace{-0.2cm}
	 	\scriptsize
	 	\begin{center}
	 		\begin{tabular}{>{\arraybackslash}m{0.8in}|>{\centering\arraybackslash}m{0.35in}|>{\centering\arraybackslash}m{0.35in}|>{\centering\arraybackslash}m{0.35in}|>{\centering\arraybackslash}m{0.35in}|>{\centering\arraybackslash}m{0.35in}|>{\centering\arraybackslash}m{0.35in}}
	 			%\cline{2-5}
	 			\multicolumn{1}{c|}{} & PESQ & CSIG & CBAK & COVL & SSNR & STOI  \parbox{0pt}{\rule{0.pt}{0ex+\baselineskip}}\\ [0ex] 	% inserts table heading
	 			\hline
	 			\hline
	 			Noisy & $1.97$ & $3.35$ & $2.44$ & $2.63$  & $1.68$ & $92.10$ \parbox{0.5pt}{\rule{0pt}{0ex+\baselineskip}}\\ [0ex] 	% inserts
	 			%SEGAN \cite{Pascual2017} & $2.16$ & $3.48$ & $2.94$ & $2.80$ & $7.73$ & $-$ \parbox{0.5pt}{\rule{0pt}{0ex+\baselineskip}}\\ [0ex] 	% inserts
	 			SEGAN \cite{Pascual2017} & $2.19$ & $3.39$ & $2.90$ & $2.76$ & $7.36$ & $93.12$ \parbox{0.5pt}{\rule{0pt}{0ex+\baselineskip}}\\ [0ex] 	% inserts
	 			\hline
	 			\hline
	 			ISEGAN \cite{Phan2020} & $2.24$ & $3.23$ & $2.95$ & $2.69$ & $8.17$ & $93.29$ \parbox{0.5pt}{\rule{0pt}{0ex+\baselineskip}}\\ [0ex] 	% inserts
	 			DSEGAN \cite{Phan2020} & $2.35$ & $3.55$ & $3.1$ & $2.93$ & $8.7$ & $93.25$ \parbox{0.5pt}{\rule{0pt}{0ex+\baselineskip}}\\ [0ex] 	% inserts
	 			\hline
	 			\hline
	 			SASEGAN-3 & $\bm{2.32}$ & $\bm{3.51}$ & $\bm{3.07}$ & $\bm{2.90}$ & $\bm{8.53}$ & $\bm{93.35}$  \parbox{0.5pt}{\rule{0pt}{0ex+\baselineskip}}\\ [0ex] 	% inserts
	 			SASEGAN-4 & $\bm{2.36}$ & $\bm{3.57}$ & $\bm{3.08}$ & $\bm{2.95}$ & $\bm{8.38}$ & $\bm{93.47}$  \parbox{0.5pt}{\rule{0pt}{0ex+\baselineskip}}\\ [0ex] 	% inserts
	 			SASEGAN-5 & $\bm{2.31}$ & $\bm{3.46}$ & $\bm{2.94}$ & $\bm{2.85}$ & $7.20$ & $\bm{93.22}$  \parbox{0.5pt}{\rule{0pt}{0ex+\baselineskip}}\\ [0ex] 	% inserts
	 			SASEGAN-6 & $\bm{2.38}$ & $\bm{3.46}$ & $\bm{3.12}$ & $\bm{2.90}$ & $\bm{8.86}$ & $\bm{93.39}$  \parbox{0.5pt}{\rule{0pt}{0ex+\baselineskip}}\\ [0ex] 	% inserts
	 			SASEGAN-7 & $\bm{2.30}$ & $\bm{3.52}$ & $\bm{2.98}$ & $\bm{2.89}$ & $7.34$ & $\bm{93.38}$  \parbox{0.5pt}{\rule{0pt}{0ex+\baselineskip}}\\ [0ex] 	% inserts
	 			SASEGAN-8 & $\bm{2.34}$ & $\bm{3.55}$ & $\bm{3.03}$ & $\bm{2.92}$ & $\bm{8.03}$ & $\bm{93.24}$ \parbox{0.5pt}{\rule{0pt}{0ex+\baselineskip}}\\ [0ex] 	% inserts
	 			SASEGAN-9 & $\bm{2.29}$ & $\bm{3.45}$ & $\bm{3.05}$ & $\bm{2.85}$ & $\bm{8.48}$ & $\bm{93.28}$  \parbox{0.5pt}{\rule{0pt}{0ex+\baselineskip}}\\ [0ex] 	% inserts
	 			SASEGAN-10 & $\bm{2.41}$ & $\bm{3.62}$ & $\bm{3.06}$ & $\bm{2.99}$ & $\bm{7.87}$ & $\bm{93.36}$  \parbox{0.5pt}{\rule{0pt}{0ex+\baselineskip}}\\ [0ex] 	% inserts
	 			SASEGAN-11 & $\bm{2.35}$ & $\bm{3.57}$ & $\bm{3.03}$ & $\bm{2.94}$ & $\bm{7.76}$ & $\bm{93.19}$ \parbox{0.5pt}{\rule{0pt}{0ex+\baselineskip}}\\ [0ex] 	% inserts
	 			Average & $\it \mathbf{2.34}$ & $\bm{3.52}$ & $\bm{3.04}$ & $\bm{2.91}$ & $\bm{8.05}$ & $\bm{93.32}$ \parbox{0.5pt}{\rule{0pt}{0ex+\baselineskip}}\\ [0ex] 	% inserts
	 			\hline
	 			SASEGAN-All & $\bm{2.36}$ & $\bm{3.54}$ & $\bm{3.08}$ & $\bm{2.93}$ & $\bm{8.27}$ & $\bm{93.49}$  \parbox{0.5pt}{\rule{0pt}{0ex+\baselineskip}}\\ [0ex] 	% inserts
	 			
	 		\end{tabular}
	 	\end{center}
	 	\label{tab:results}
	 	\vspace{-0.4cm}
	 \end{table}

	\begin{figure} [!t]
	\centering
	\includegraphics[width=0.8\linewidth]{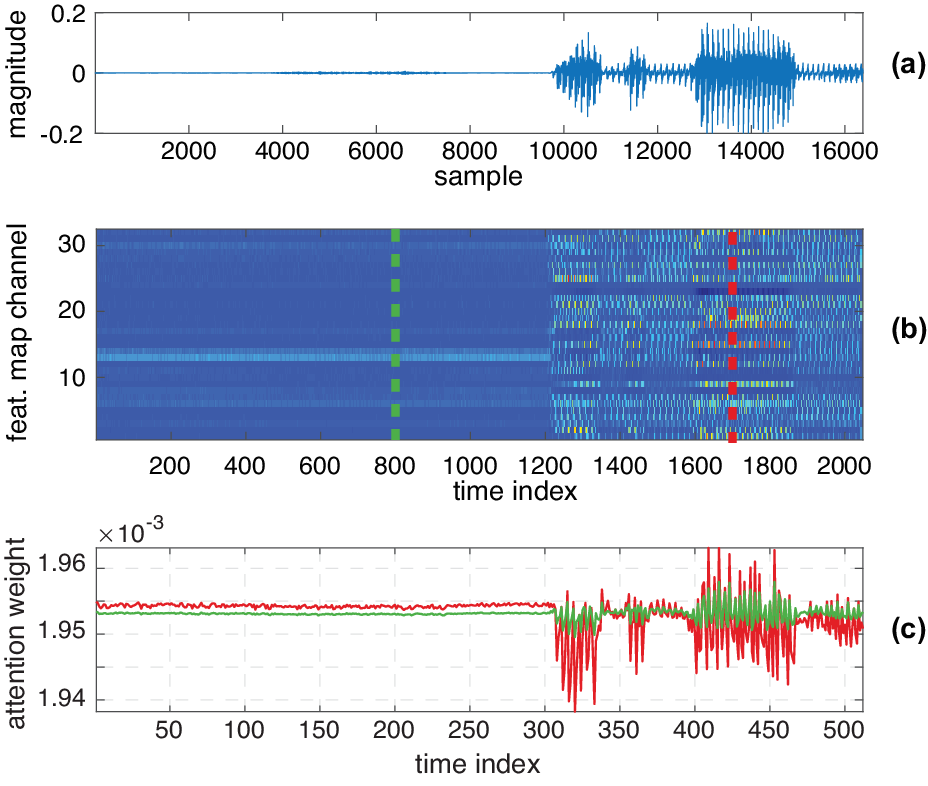}
	\vspace{-0.4cm}
	\caption{\small Visualization of self-attention weights at the $3^{rd}$ convolutional layer of the generator's encoder. (a) the raw speech input; (b) the feature map of size $2048 \times 32$; and (c) the attention weights distributed over $512$ time indices of the matrix $\mathbf{V}$ (note that the factor $p=4$). The green and red distributions in (c) correspond to the two locations specified by the green and red dash lines in (b).}
	\label{fig:attention_weight}
	\vspace{-0.25cm}
\end{figure}
	
	\vspace{-0.25cm}
	\subsection{Discussion}
	\vspace{-0.25cm}
	
	In order visualize the learned attention weights, taking the case SASEGAN-$3$ for example, we exhibit in Fig.~\ref{fig:attention_weight} those attention weights of the generator's encoder corresponding to different temporal locations of the feature map. This suggests that the network leverages complementary features in distant portions of the input rather than local regions of fixed shape to generate the attentive output. Apparently, more attention was put on speech regions (the red distribution in Fig.~\ref{fig:attention_weight} (c)) to synthesize the new feature at the location with speech (at the red dashed line in Fig.~\ref{fig:attention_weight} (b)). The opposite is observed for the green distribution in Fig.~\ref{fig:attention_weight} (c) for the location in background noise region (the green dashed line in Fig.~\ref{fig:attention_weight} (b)).
	
	It is also worth mentioning other improved SEGANs \cite{Phan2020,Zhang2020,Baby2019} that have been built upon the first SEGAN in \cite{Pascual2017}. While the proposed SASEGAN performs competitively to the existing SEGAN variants, for example ISEGAN and DSEGAN \cite{Phan2020} (cf.~Table~\ref{tab:results}), our primary goal in this work is to study the influence of the proposed self-attention layer and its placement on speech enhancement rather than a comprehensive comparison among SEGAN variants. More importantly, the proposed self-attention layer is generic enough that it can be  applied to those existing SEGAN variants to further improve their performance. We leave this for a future study.
	
	\vspace{-0.3cm}
	\section{Conclusions}
	\vspace{-0.3cm}	
	We proposed and integrated a self-attention layer with SEGAN to improve its temporal dependency modeling for speech enhancement. The proposed self-attention layer can be used at different (de)convolutional layers of the SEGAN's generator and discriminator or even all of them, given sufficient processing memory. Our experiments show that the self-attention SEGAN outperforms the SEGAN baseline over all of the objective evaluation metrics. In addition, consistency in the improvement was seen across the self-attention placement settings. Furthermore, these settings did not result in a significant difference among their performance gains. The results suggest that self-attention can be sufficiently used in a high-level (de)convolutional layer with very small induced memory. Furthermore, it can be easily applied to existing SEGAN variants for potential improvement.
	
	%\pagebreak
	%\newpage
	%\footnotesize
	\small
	\bibliographystyle{IEEEbib}
	\bibliography{refgan}

\begin{thebibliography}{10}

\bibitem{Donahue2018}
C.~Donahue, B.~Li, and R.~Prabhavalkar,
\newblock ``Exploring speech enhancement with generative adversarial networks
  for robust speech recognition,''
\newblock in {\em Proc. ICASSP,}, 2018, pp. 5024--5028.

\bibitem{Xu2019}
Y.~Xu \emph{et al.},
\newblock ``Joint training of complex ratio mask based beamformer and acoustic
  model for noise robust {ASR},''
\newblock in {\em Proc. ICASSP}, 2019, pp. 6745--6749.

\bibitem{Weninger2015}
F.~Weninger \emph{et al.},
\newblock ``Speech enhancement with {LSTM} recurrent neural networks and its
  application to noise-robust asr,''
\newblock {\em Proc. Intl. Conf. on Latent Variable Analysis and Signal
  Separation}, pp. 91--99, 2015.

\bibitem{Schasse2014}
A.~Schasse \emph{et al.},
\newblock ``Two-stage filter-bank system for improved single-channel noise
  reduction in hearing aids,''
\newblock {\em IEEE/ACM Transactions on Audio, Speech, and Language
  Processing}, vol. 23, no. 2, pp. 383--393, 2014.

\bibitem{Yang2005}
L.-P. Yang and Q.-J. Fu,
\newblock ``Spectral subtraction-based speech enhancementfor cochlear implant
  patients in background noise,''
\newblock {\em Journal of the Acoustical Society of America}, vol. 117, no. 3,
  pp. 1001--1004, 2005.

\bibitem{Gerkmann2011}
T.~Gerkmann and R.~C. Hendriks,
\newblock ``Unbiased {MMSE}-based noise power estimation with low complexity
  and low tracking delay,''
\newblock {\em IEEE Trans. on Audio, Speech, and Language Processing}, pp.
  1383--1393, 2011.

\bibitem{Ephraim1985}
Y.~Ephraim and D.~Malah,
\newblock ``Speech enhancement using a minimum mean-square error log-spectral
  amplitude estimator,''
\newblock {\em IEEE Trans. on Acoustics, Speech, and Signal Processing}, vol.
  33, no. 2, pp. 443--445, 1985.

\bibitem{Xu2015}
Y.~Xu \emph{et al.},
\newblock ``A regression approach to speech enhancement based on deep neural
  networks,''
\newblock {\em IEEE/ACM Trans. on Audio, Speech and Language Processing
  (TASLP)}, vol. 23, no. 1, pp. 7--19, 2015.

\bibitem{Li2018}
Z.~X.~Li \emph{et al.},
\newblock ``A conditional generative model for speech enhancement,''
\newblock {\em Circuits, Systems, and Signal Processing}, vol. 37, no. 11, pp.
  5005--5022, 2018.

\bibitem{Park2017}
S.~R. Park and J.~Lee,
\newblock ``A fully convolutional neural network for speech enhancement,''
\newblock in {\em Proc. Interspeech}, 2017.

\bibitem{Erdogan2015}
H.~Erdogan \emph{et al.},
\newblock ``Phase sensitive and recognition-boosted speech separation using
  deep recurrent neural networks,''
\newblock in {\em Proc. ICASSP}, 2015, pp. 708--712.

\bibitem{Goodfellow2014}
I.~Goodfellow \emph{et al.},
\newblock ``Generative adversarial nets,''
\newblock in {\em Proc. NIPS}, 2014, pp. 2672--2680.

\bibitem{Pascual2017}
S.~Pascual, A.~Bonafonte, and J.~Serr\`{a},
\newblock ``{SEGAN}: Speech enhancement generative adversarial network,''
\newblock in {\em Proc. Interspeech}, 2017, pp. 3642--3646.

\bibitem{Zhang2020}
Z.~Zhang \emph{et al.},
\newblock ``On loss functions and recurrency training for {GAN}-based speech
  enhancement systems,''
\newblock in {\em Proc. Interspeech}, 2020.

\bibitem{Phan2020}
H.~Phan \emph{et al.},
\newblock ``Improving {GAN}s for speech enhancement,''
\newblock {\em IEEE Signal Processing Letters}, vol. 27, pp. 1700--1704, 2020.

\bibitem{Baby2019}
D.~Baby and S.~Verhulst,
\newblock ``{SERGAN}: Speech enhancement using relativistic generative
  adversarial networks with gradient penalty,''
\newblock in {\em Proc. ICASSP}, 2019, pp. 106--110.

\bibitem{Defossez2020}
A.~Defossez, G.~Synnaeve, and Y.~Adi,
\newblock ``Real time speech enhancement in the waveform domain,''
\newblock in {\em Proc. Interspeech}, 2020.

\bibitem{Pham2019}
N.-Q.~Pham \emph{et al.},
\newblock ``Very deep self-attention networks for end-to-end speech
  recognition,''
\newblock in {\em Proc. Interspeech}, 2019.

\bibitem{Sperber2018}
M.~Sperber \emph{et al.},
\newblock ``Self-attentional acoustic models,''
\newblock in {\em Proc. Interspeech}, 2018, pp. 3723--3727.

\bibitem{Tian2019}
Z.~Tian \emph{et al.},
\newblock ``Self-attention transducers for end-to-end speech recognition,''
\newblock in {\em Proc. Interspeech}, 2019, pp. 4395--4399.

\bibitem{Wang2018}
X.~Wang \emph{et al.},
\newblock ``Non-local neural networks,''
\newblock in {\em Proc. CVPR}, 2018.

\bibitem{Zhang2019}
H.~Zhang \emph{et al.},
\newblock ``Self-attention generative adversarial networks,''
\newblock in {\em Proc. ICML}, 2019, pp. 7354--7363.

\bibitem{Mao2017}
X.~Mao \emph{et al.},
\newblock ``Least squares generative adversarial networks,''
\newblock in {\em Proc. ICCV}, 2017, pp. 2813--2821.

\bibitem{Radford2016}
A.~Radford, L.~Metz, and S.~Chintala,
\newblock ``Unsupervised representation learning with deep convolutional
  generative adversarial networks,''
\newblock in {\em Proc. ICLR}, 2016.

\bibitem{He2015}
K.~He \emph{et al.},
\newblock ``Delving deep into rectifiers: Surpassing human-level performance on
  imagenet classification,''
\newblock in {\em Proc. ICCV}, 2015, pp. 1026--1034.

\bibitem{Salimans2016}
T.~Salimans \emph{et al.},
\newblock ``Improved techniques for training {GAN}s,''
\newblock in {\em Proc. NIPS}, 2016, pp. 2226--2234.

\bibitem{Maas2013}
A.~L. Maas, A.~Y. Awni, and A.~Y. Ng,
\newblock ``Rectifier nonlinearities improve neural network acoustic models,''
\newblock in {\em Proc. ICML}, 2013, vol.~30.

\bibitem{Miyato2018}
T.~Miyato \emph{et al.},
\newblock ``Spectral normalization for generative adversarial networks,''
\newblock in {\em Proc. ICLR}, 2018.

\bibitem{Botinhao2016}
C.~Valentini-Botinhao \emph{et al.},
\newblock ``Investigating {RNN}-based speech enhancement methods for
  noise-robust text-to-speech,''
\newblock in {\em Proc. 9th ISCA Speech Synthesis Workshop}, 2016, pp.
  146--152.

\bibitem{Veaux2013}
C.~Veaux, J.~Yamagishi, and S.~King,
\newblock ``The voice bank corpus: design, collection and data analysis of a
  large regional accent speech database,''
\newblock in {\em Proc. 2013 International Conference Oriental COCOSDA}, 2013,
  pp. 1--4.

\bibitem{Thiemann2013}
J.~Thiemann, N.~Ito, and E.~Vincent,
\newblock ``The diverse environments multi-channel acoustic noise database: A
  database of multichannel environmental noise recordings,''
\newblock {\em The Journal of the Acoustical Society of America}, vol. 133, no.
  5, pp. 3591--3591, 2013.

\bibitem{Abadi2016}
M.~{Abadi \emph{et al.}},
\newblock ``Tensorflow: Large-scale machine learning on heterogeneous
  distributed systems,''
\newblock in {\em Proc. 12th {USENIX} symposium on operating systems design and
  implementation ({OSDI} 16)}, 2016, pp. 265--283.

\bibitem{Tieleman2012}
T.~Tieleman and G.~Hinton,
\newblock ``Lecture 6.5 - {RMSprop}: divide the gradient by a running average
  of its recent magnitude,''
\newblock {\em Coursera: Neural Networks for Machine Learning}, 2012.

\bibitem{Taal2011}
H.~Taal \emph{et al.},
\newblock ``An algorithm for intelligibility prediction of time-frequency
  weighted noisy speech,''
\newblock {\em IEEE Trans. Speech Audio Processing}, vol. 19, no. 7, pp.
  2125–2136, 2011.

\end{thebibliography}
\end{document}